# Hard x-ray standing-wave photoemission insights into the structure of an epitaxial Fe/MgO multilayer magnetic tunnel junction

**Authors:** C.S. Conlon[1,2], G. Conti[1,2], S. Nemšák[3], G. Palsson[4], R. Moubah[4], C.-T. Kuo[1,2], M. Gehlmann[1,2], J. Ciston[5], J. Rault[6], J.-P. Rueff[6,7], F. Salmassi[8], W. Stolte[3], A. Rattanachata[1,2], S.-C. Lin[1,2], A. Keqi[1,2], A. Saw[1,2], B. Hjörvarsson[4], and C.S. Fadley[1,2]

**Affiliations:**

1 Department of Physics, University of California, Davis, Davis, CA, USA 95616;

2 Materials Sciences Division, Lawrence Berkeley National Laboratory, Berkeley, CA, USA 94720;

3 Advanced Light Source, Lawrence Berkeley National Laboratory, Berkeley, CA, USA 94720;

4 Department of Physics, 752 37 Uppsala University, Uppsala, Sweden;

5 National Center for Electron Microscopy, Molecular Foundry, Lawrence Berkeley National Laboratory, Berkeley, CA, USA 94720;

6 SOLEIL Synchrotron, 91190 Saint-Aubin, France;

7 Sorbonne Université, CNRS, Laboratoire de Chimie Physique-Matière et Rayonnement, LCPMR, F-75005, Paris, France;

8 Center for X-ray Optics, Lawrence Berkeley National Laboratory, Berkeley, CA, USA 94720

**ABSTRACT:**

The Fe/MgO magnetic tunnel junction is a classic spintronic system, with current importance technologically, and interest for future innovation. The key magnetic properties are linked directly to the structure of hard-to-access buried interfaces, and the Fe and MgO components near the surface are unstable when exposed to air, making a deeper probing, non-destructive, *in-situ* measurement ideal for this system. We have thus applied hard x-ray photoemission spectroscopy (HXPS) and standing-wave (SW) HXPS in the few keV energy range to probe the structure of an epitaxially-grown MgO/Fe superlattice. The superlattice consists of 9 repeats of MgO grown on Fe by magnetron sputtering on an MgO (001) substrate, with a protective $Al_2O_3$ capping layer. We determine through SW-HXPS that 8 of the 9 repeats are similar and ordered, with a period of 33 ± 4 Å, with minor presence of FeO at the interfaces and a significantly distorted top bilayer with ca. 3 times the oxidation of the lower layers at the top MgO/Fe interface. There is evidence of asymmetrical oxidation on the top and bottom of the Fe layers. We find agreement with dark-field scanning transmission electron microscope (STEM) and x-ray reflectivity measurements. Through the STEM measurements we confirm an overall epitaxial stack with dislocations and warping at the interfaces of ca. 5 Å. We also note a distinct difference in the top bilayer, especially MgO, with possible Fe inclusions. We thus demonstrate that SW-HXPS can be used to probe deep buried interfaces of novel magnetic devices with few-angstrom precision.

INTRODUCTION:

Since they were first realized in the late 1980s, multilayer magnetic heterostructures showing the giant magneto-resistance (GMR) effect have led to great leaps in the understanding of coupled magnetic systems and magnetic data storage technology [1, 2, 3]. Metallic GMR structures were then augmented by multilayer oxide/metal magnetic tunnel junctions (MTJ's) that are now ubiquitous in spintronic devices such as read heads and magnetic random access memory (MRAM) [4, 5, 6, 7, 8, 9, 10]. Intense efforts have been devoted to magnetic nanostructures to explore perpendicular magnetic anisotropy, novel geometries and optimal dopant materials at the heterostructure interfaces, and to high quality epitaxial superlattices to create smaller, faster, and more energy efficient spintronic devices. One classical system is the Fe/MgO/Fe MTJ, with recent developments proving additions of new dopants [11], novel structures [12], or new growth techniques [13] can result in tunneling magnetoresistance (TMR) ratios of >600% [11] or emergent magnetic properties such as layer-by-layer magnetic switching in multilayer superlattices [13].

Over the decades of study of the Fe/MgO/Fe system, a few key structural components have been realized to be crucial to TMR and magnetic properties of the MTJ, including layer thickness [14], atomic order [11, 15, 16], and oxygen concentration [9, 10, 11, 12, 14]. Measurements of TMR variation based on the MgO layer thickness show a strong variation in the first few layers of MgO, which then levels off for greater thicknesses [9, 10]. Certain mechanisms have been proposed for the Fe layer thickness (and associated Fe lattice relaxation) dependent magnetization, with thicker Fe layers resulting in higher interfacial Fe magnetic moments up to 11 Fe monolayers [14]. Oxygen vacancies in MgO at the interface could lead to lower TMR [17], and the presence of FeO at the interface has been measured to coincide with decreased magnetization at the interface which can also result in decreased TMR [18]. Theoretical calculations propose that increasing oxygen concentration at the interface can affect the magnetic interlayer exchange coupling in Fe/MgO/Fe MTJs, with interfacial oxygen vacancies resulting in strongest antiferromagnetic exchange, and increased oxidation suppressing this exchange and even flipping it to ferromagnetic [19]. The oxidation at the interface is influenced by interface roughness, since the Fe lattice sites are predominantly located on the O sites of MgO for Fe/MgO growth [20]. A recent study measured TMR as a function of roughness of one of the Fe/MgO interfaces by varying the coverage of an atomically flat area of Fe with monatomic Fe islands, and found the largest TMR associated with the most Fe steps [15]. Calculations of TMR on the Fe/MgO/Fe MTJ system also show symmetry of oxidation about the top and bottom MgO interface to be important, with extreme asymmetry of only one oxidized interface resulting in reduced TMR, and symmetric oxidation enhanced TMR [21]. One theme among the current literature is that the interface where Fe meets MgO is critical. To build a clear model of the magnetism of novel Fe/MgO/Fe devices an in-depth understanding of the physical structure is key.

One recent type of structure of interest is an epitaxially grown superlattice of repeated [MgO/Fe] bilayers grown on MgO(001) developed at the University of Uppsala [13]. Neutron scattering measurements have shown that in this multilayer sample the Fe layers switch magnetization layer-by-layer in an applied field, with a rotation of 90° in the orientation of the magnetization between adjacent Fe layers [13]. This interlayer coupling is not completely understood [13]. While this growth has demonstrated epitaxial order and a well-defined heterostructure [13, 22], it is critical to obtain a complete understanding of the microstructure at the heterostructure interfaces, with a method that simultaneously probes local charge and electron configurations. Two such methods are noted here. Previous studies using magnetic circular dichroism (MCD) in soft x-ray SW-XPS on an Fe/MgO MTJ demonstrate the possibilities of this technique for deriving depth dependent magnetizations [18]. Another SW-HXPS study of the depth distributions of boron in a Ta/Co$_{0.2}$Fe$_{0.6}$B$_{0.2}$/MgO multilayer [23] has further demonstrated the capabilities of this technique for studying depth-resolved properties, while additional SW-XPS studies of oxide heterostructures show possible future extensions of the SW method which incorporate x-ray photoelectron diffraction and angle-resolved photoemission spectroscopy [24]. A recent study using electron magnetic circular dichroism (EMCD) in electron energy-loss spectroscopy (EELS) on a single layer of Fe deposited on MgO (001) shows another method which combines structural and magnetic measurements near the interface [25]. Here we make use of the former, SW-XPS technique, which has the benefit of not requiring any destructive sample preparation or cleaning, and we apply it for the first time to a highly epitaxial Fe/MgO superlattice.

The standard XPS measurement is surface sensitive, where the inelastic mean free path (IMFP) of the photoelectron limits the probing depth to a few atomic layers for soft x-rays at energies of hundreds of eV, to nanometers in the tender and hard x-ray regions of thousands of eV [26]. The SW-XPS method uses the interference between an incident and reflected x-ray beam to add depth resolution and probe buried interfaces of samples [27]. We have furthermore used more energetic hard/tender x-rays in the few-keV regime to penetrate more deeply into the structure. The necessary reflectivity is established with a superlattice of alternating planar materials with different refractive indices. With x-ray incidence ($\theta_i$) at the mirror's Bragg angle ($\theta_{Bragg}$), as shown in eq. (1) and Figure 1, this is the standard rocking curve (RC) method [24] or with a sample at glancing incidence angles approaching zero, the total reflection (TR) RC method [28]. Tuning to the Bragg condition for the multilayer

$$\lambda_x = 2d_{ML}\sin(\theta_{Bragg}), \quad (1)$$

where $\lambda_x$ is the x-ray beam wavelength, and $d_{ML}$ is the superlattice period, yields a strong reflected wave. The x-ray interference in the case of a superlattice sample forms a SW with alternating high and low electric field amplitudes vertically through the sample. The period of this SW matches $d_{ML}$ of the sample. The phase of this SW can be scanned vertically through the sample by either rotating the sample ($\theta_i$) around $\theta_{Bragg}$ (the RC method), or tuning the x-ray beam energy ($\lambda_x$) with the x-ray incidence angle fixed. Both of these methods scan the SW pattern of the electric field vertically through the sample, and the RC method where the incidence angle is changed is used in this study. The peak-integrated photoemission intensity, which is weighted by this electric field profile, is measured for each $\theta_i$ angle, generating RCs for various XPS core levels.

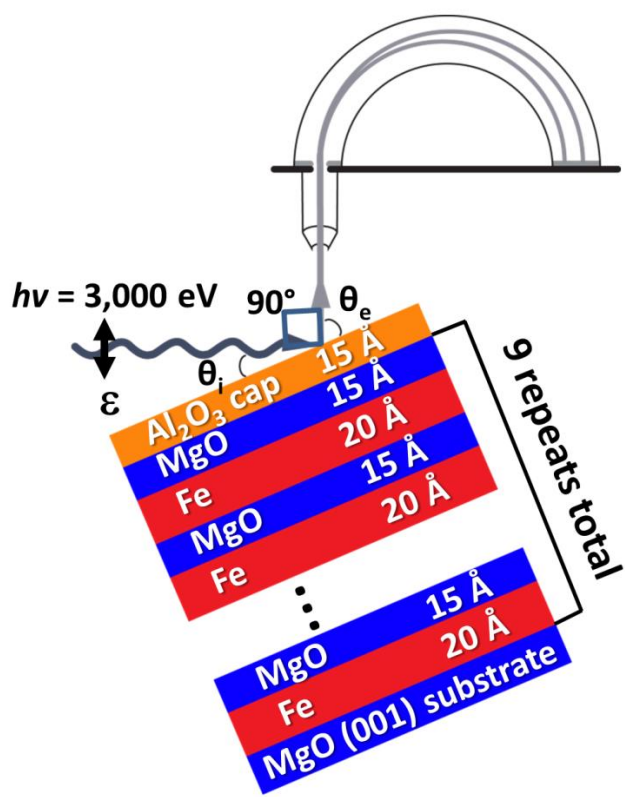

**Figure 1:** Sample with model growth parameters. Geometry for HXPS measurements is shown. The incidence angle $\theta_i$ is varied for each HXPS measurement to create the RCs, with 90° between the x-ray beamline and analyzer fixed. $\theta_e$ and $\varepsilon$ refer to the angle of electron emission and the x-ray polarization direction, respectively.

While this SW RC method allows buried interfaces to dominate the XPS measurements when the x-ray SW is in a position where the sample surface has very low electric field amplitude, the IMFP of the photoelectrons still constrains the effective probing depth of the sample. For samples with deep buried interfaces, using hard x-rays (HXPS) can extend the total depth of the measurement and increase counting statistics for deep layers. It is in the hard/tender x-ray energy regime that we have worked.

In this study we have determined the detailed structure of an epitaxial MgO/Fe superlattice grown after the method of R. Moubah et al. at University of Uppsala [13]. We have used SW-HXPS, a technique proven to non-destructively probe buried solid-solid interfaces *in-situ* [18, 23], to determine the chemical structure, oxidation state, and interdiffusion/roughness of this interface with few Å resolution. Real space imaging was also investigated by high angular dark field scanning tunneling electron microscopy (STEM). Our results provide a detailed understanding of the structure, and specifically the interface, which is critical in being able to model emergent magnetic properties.

RESULTS AND DISCUSSION:

A superlattice of nominally [MgO (15 Å)/Fe (20 Å)]$_9$ was grown by magnetron sputtering on an MgO (001) substrate in an Ar atmosphere by the B. Hjörvarsson group at Uppsala University. The nominal structure of the sample as grown is shown in Figure 1, and more details on sample growth are presented in R. Moubah et al. and H. Raanaei et al. [13, 22]. A 15 Å Al$_2$O$_3$ capping layer was grown on top to protect the top MgO surface, as MgO is sensitive to air and x-ray exposure. All measurements presented below were performed on this sample.

HXPS data from the multilayer sample were obtained at bend-magnet beamline 9.3.1 at the Advanced Light Source (ALS) of the Lawrence Berkeley National Lab (LBNL), utilizing a Scienta SES 2002 spectrometer equipped with a five-axis specimen manipulator/goniometer. At 3,000 eV the total beam and spectrometer experimental resolution is 0.6 eV. The experimental geometry is shown in Figure 1, with $\theta_i$ used to indicate the incidence angle between the x-ray beam and the sample surface. The radiation polarization $\varepsilon$ lies in the photoemission plane and is thus p-type. The angle between the x-ray beam and spectrometer is fixed at 90°. The sample HXPS survey and example core levels were all taken at normal emission of the photoelectrons ($\theta_e$ = 90°) and show no contamination of the sample except for the usual adsorbed C and O on the surface, as shown by the large-scale scans in Figure 2-3. The coverage of these contaminants is estimated from C 1s and O 1s core level intensities and comparison to simulations of electron spectra using the simulation of electron spectra for surface analysis (SESSA) program [29] to be 10 Å at a ratio of 0.8 O:C. The estimation with the SESSA program uses the HXPS peak intensities and is a rough estimation which requires other measurements for refinements, including the detailed depth distribution of species. Comparing multiple measurements at additional $\theta_i$, such as in the SW-XPS RC method, or at additional x-ray energies (not done in this study) will result in a more quantitative estimate. The SW-XPS RC method we use yields a much more refined structure since it uses dozens of HXPS measurements, each with distinct electric field profiles, in the structure determination. Table I shows the normalized XPS peak areas of the experimental core level spectra (Figure 3 and 4) and the SESSA simulation of the sample geometry which best fit these values. The basic model we used in the SESSA program is discussed later in comparison with the more complex model used for the SW-XPS RCs. To determine the number of O 1s peaks used to fit the O 1s core level spectrum shown in Figure 3, the SW-XPS RC data was used to ensure each of the 3 peaks had a distinct RC phase, as we describe later in more detail.

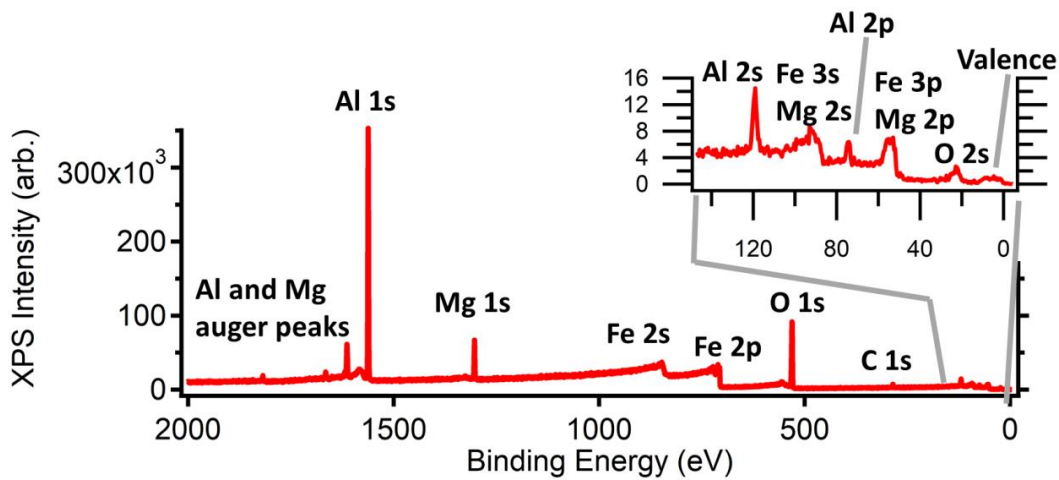

**Figure 2:** HXPS survey data at 3,000 eV excitation energy. No contamination peaks are present except expected carbon peak from surface contamination due to air exposure, and some oxygen beyond that directly from the Fe/MgO sample itself.

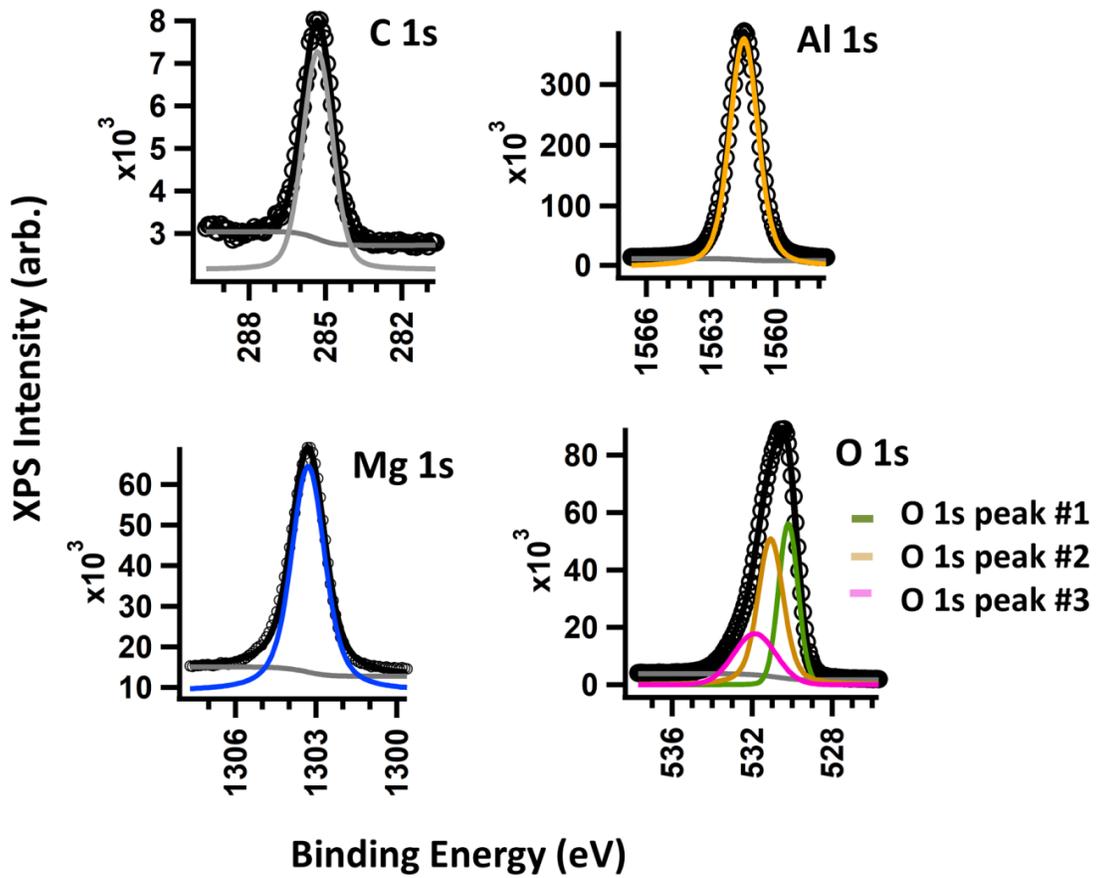

**Figure 3:** HXPS data at 3,000 eV excitation energy. Shown are experimental data for core levels (open circles) averaged over all RC spectra with example peak fits (Voigt functions, colored curves) and Shirley backgrounds (gray curves). See figure 4 for Fe 2p core level.

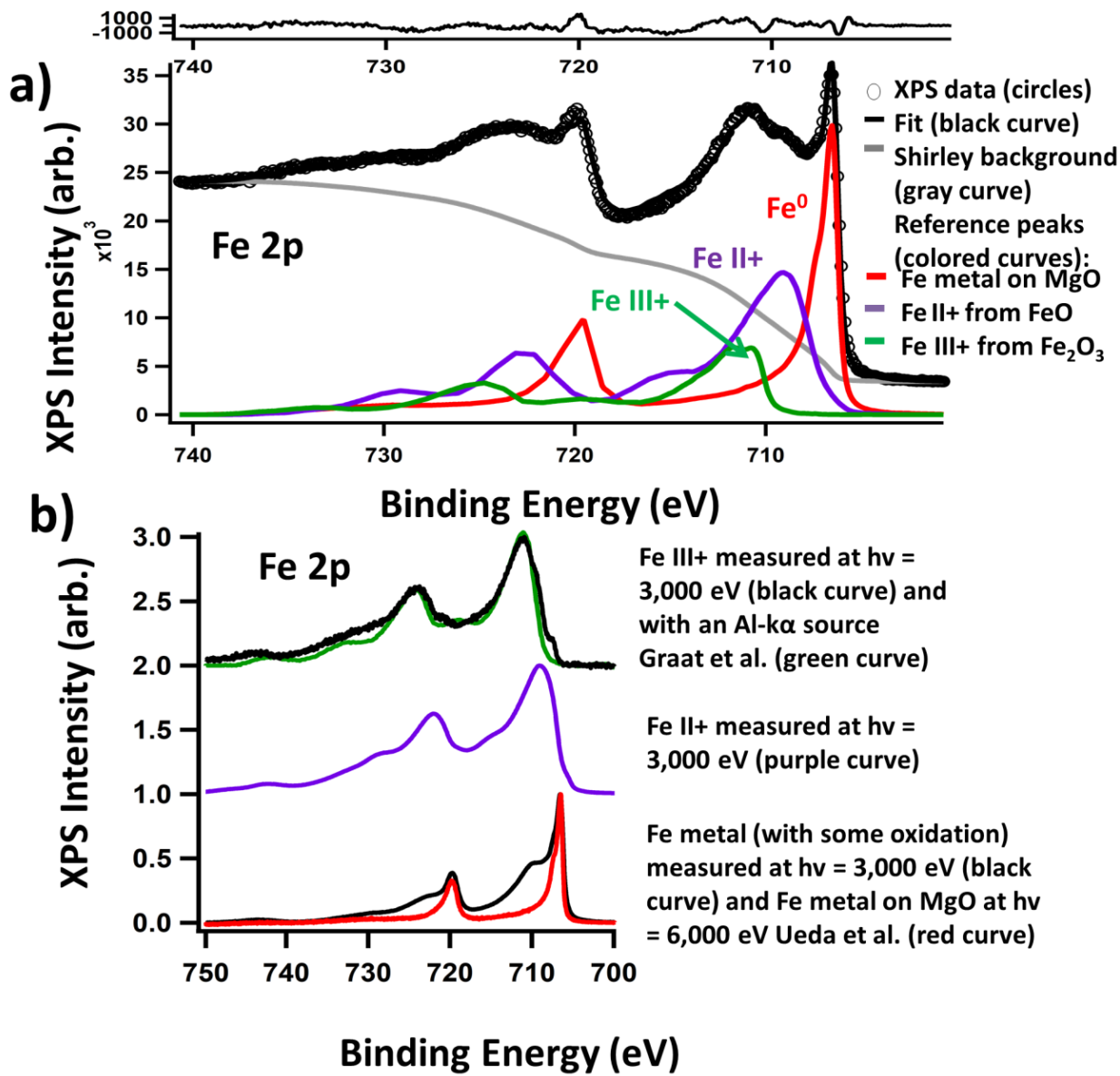

**Figure 4: a)** HXPS data at 3,000 eV excitation energy for the Fe 2p core level, averaged over all RC spectra (open circles). This spectrum was fitted (black curve) using reference spectra for metallic Fe from 20 nm Fe deposited on MgO at 6,000 eV from Ueda et al. [34] broadened with a Gaussian convolution to match experimental resolutions (red curve), Fe II+ from pressed FeO powder at 3,000 eV (purple curve), and Fe III+ from pressed $Fe_2O_3$ powder with an Al-kα source from Graat et al. [35] (Green curve). The residual from the fit is shown at the top. All reference spectra and sample spectrum have Shirley background removed (shown in gray for sample spectrum). **b)** Spectra in black and purple are reference samples from (bottom to top) Fe foil, FeO pressed powder, and $Fe_2O_3$ pressed powder measured at 3,000 eV. In red, purple, and green are reference curves used in figure 4a for metallic Fe, Fe II+ and Fe III+, respectively. References are vertically offset for ease of comparison.

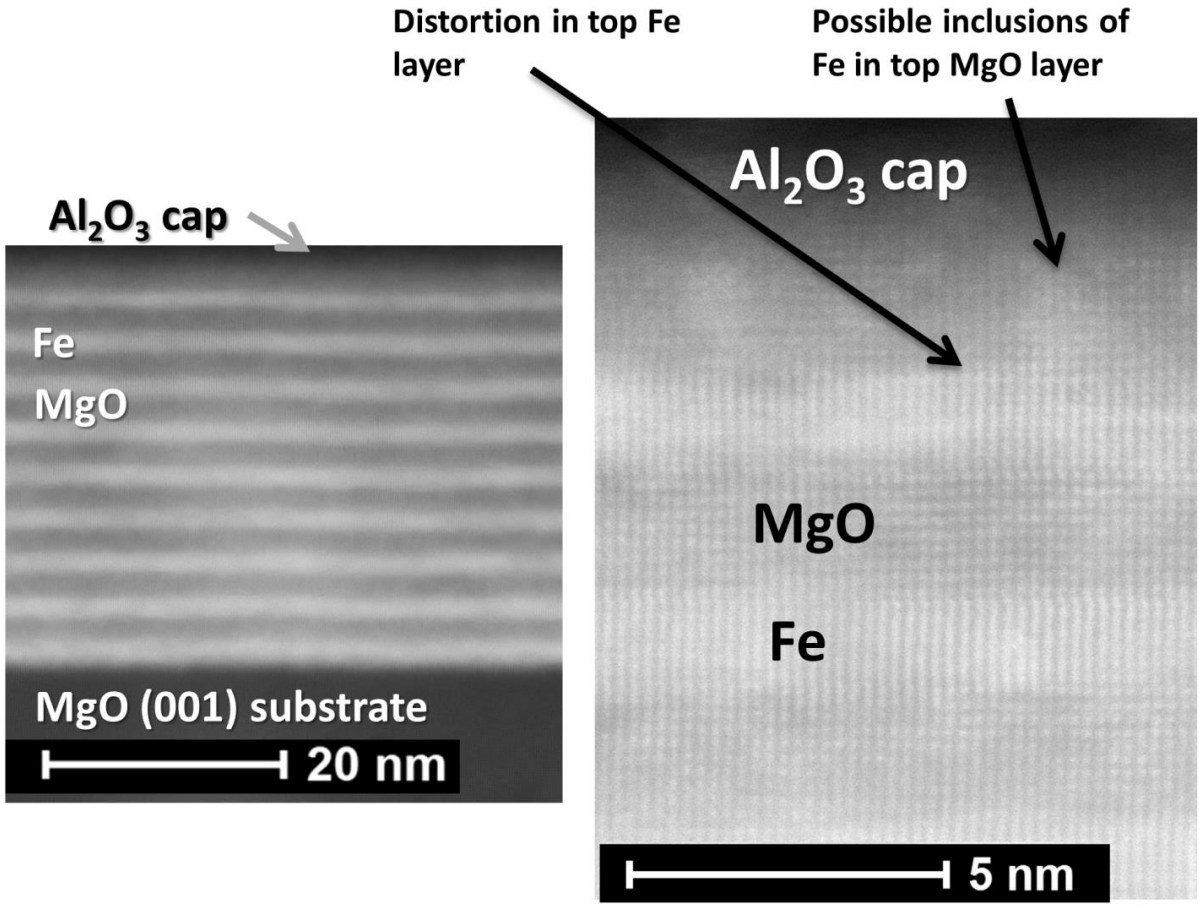

**Figure 5**: Dark field STEM for the sample, showing full sample stack (left) and detailed scan of top layers (right). Pt/C cap (not shown) was added during sample preparation to protect the sample.

STEM measurements (Figure 5) were performed at the TEAM 0.5 microscope at the National Center for Electron Microscopy facility of the Molecular Foundry at LBNL, operated at 300 kV. Geometric aberrations were corrected to third order with a 17 mrad convergence semiangle. A cross-section STEM sample was removed from the center of the sample used for the HXPS measurements, in the measurement region for the HXPS, and prepared with the Focused Ion Beam lift out method. A Pt/C protective cap was deposited for this process and the sample preparation was performed shortly before imaging to minimize total oxidation of the Fe layers after milling.

**Table I:** Integrated fitted XPS peak area for various core levels at 3,000 eV. All values are normalized within their column to the peak area of the Mg 1s core level.

| XPS core level | Average total XPS intensity from RC (normalized to Mg 1s intensity) | Average total XPS intensity from SESSA (normalized to Mg 1s intensity) |
|---|---|---|
| C 1s | 0.08 | 0.08 |
| O 1s peak # 3 | 0.2 | 0.25 |
| Al 1s | 6.72 | 6.5 |
| O 1s peak #2 | 1.24 | 2.02 |
| Mg 1s | 1 | 1 |
| O 1s peak #1 | 0.63 | 0.74 |
| Fe II+ 2p | 1.27 | 1.31 |
| Fe metal 2p | 1.08 | 1.17 |
| Total Fe 2p | 2.79 (includes some Fe III+ in experimental fit) | 2.48 (No Fe III+ included in model) |

We have used XPS in the hard x-ray energy regime (HXPS) at 3,000 eV in order to more deeply probe into the sample, and access both the top MgO/Fe interface, and the Fe/MgO interface below that, expected ~30 and 50 Å below the sample surface (cf. Figure 1). Hard x-ray energies are necessitated by both the relatively short IMFP of Fe photoelectrons and the depth of the primary interface of interest. An excitation energy of 3,000 eV results in a photoelectron IMFP ($\Lambda_e$) of ~40 Å for Fe 2p as calculated by the TPP-2M formula [30]. The exponential probing depth of this experiment can be approximated by considering the attenuation of the excited photoelectrons as they move through the sample layers above the emitting atom, according to [31],

$$I(d) = I_0 \exp\left(-\frac{d}{\Lambda(E)\sin(\theta_e)}\right), \qquad (2)$$

where *I(d)* is the intensity of photoelectrons excited at depth *d* with initial intensity $I_0$, $\Lambda(E)$ is the IMFP of the photoelectron of interest at photoelectron kinetic energy *E*, and $\theta_e$ is the electron exit angle relative to the surface. Our analysis of the SW data will permit determining the various layer and interface compositions and thicknesses.

Since the IMFP differs for the different chemical species of atom, the initial core level of the photoelectron, and the properties of the sample layer that the photoelectron is traversing, we use a distinct exponential for each layer the photoelectron must travel through to reach the surface. The resulting equation, with intensity integrated and summed over layers of type *i*, and inelastic scattering in layers *j* above is:

$$I_i = I_0 \sum_{i=1}[1 - \exp(-d_i/\Lambda_{e,i}\sin\theta_e)] \prod_{j\ above\ i} \exp(-d_j/\Lambda_{e,j}\sin\theta_e), \qquad (3)$$

with $d_i$ or $d_j$ the respective thicknesses of different layers through the sample, $\Lambda_{e,i}$ or $\Lambda_{e,j}$ the respective inelastic mean free paths in the different layers, and $\theta_e$ again the electron emission angle with respect to the surface. Setting initial intensity $I_o$ = 1 results in the fraction of photoelectrons exiting a sample from a specific type of layer (e.g. Fe or MgO) and orbital (e.g. Fe 2p or Mg 1s) of interest. We use the SESSA database for IMFP values [29]. For example, for Fe $2p_{3/2}$ electrons originating in a metallic Fe layer, we expect 12% of these emitted electrons at a sample depth of 80 Å to reach the sample surface. Thus, we can in effect measure the first two bilayers of MgO/Fe in the multilayer sample.

Voigt fits after Shirley background removal were used to determine peak area for all core level peaks except in the case of the Fe 2p spectra. Due to the overlapping of many photoemission satellite peaks in the metallic and oxide spectra for Fe, the reference Fe 2p spectra described below were used to determine lineshapes for three different chemical species of Fe, which were used to fit the Fe 2p spectra (Figure 4).

Reference spectra measured at the GALAXIES beamline at SOLEIL, Saint-Aubin, France [32], replicate conditions as closely as possible of the rocking curve HXPS study of the multilayer sample at the ALS beamline 9.3.1 [33]. Commercially purchased Fe metal foil (99.994 % pure), FeO powder (99.8% pure), and $Fe_2O_3$ powder (99.995% pure) for Fe 2p lineshapes of Fe metal, Fe II+, and Fe III+, respectively, were measured with overall resolution, experimental geometry, and photon excitation energy matching the HXPS performed at the ALS on the multilayer sample. These reference samples were measured at the GALAXIES beamline using its U20 undulator, Scienta EW4000 spectrometer, and five-axis manipulator [32]. All measurements were at $\theta_e = 90°$ except for the $Fe_2O_3$ pellet which was measured at $\theta_e = 45°$ to reduce sample charging compared to normal emission, for which the photon flux is a maximum [31]. Due to these charging effects on $Fe_2O_3$ and to ensure that no surface oxidation from the Fe foil affected the final fits, spectra from the literature were used as well (Figure 4a and b). Shown in Figure 3 and Figure 4a are lineshapes used for fitting the rocking curves: lineshapes for Fe metal from S. Ueda [34], for Fe II+ from GALAXIES (see above), and for Fe III+ from P. Graat et al. [35]. Lineshapes agreed across all reference spectra, with minor differences in elastic scattering from differences in excitation energy and differences in experimental resolution well below experimental error. See Figure 4b for a comparison of the lineshapes used to those measured at GALAXIES. From this figure we can see that after Shirley background subtraction the lineshapes are in good agreement, except for minor oxidation and charging effects. Because of the oxidation of the metallic Fe sample and the charging of the Fe III+ sample at GALAXIES, the substitution of lineshapes from literature [34, 35] were made for the final fitting. In the case of the metallic Fe spectrum from Ueda et al. [34] a Gaussian convolution was applied to this higher resolution (0.15 eV) literature spectrum to match the experimental resolution of the sample (0.6 eV). All spectra from the references had Shirley backgrounds removed prior to fitting.

We have measured RCs by varying the x-ray incidence angle, $\theta_i$, over the Bragg peak of the multilayer, and also measured intensities in the near-zero TR range as a complementary SW method [28]. The Bragg peak is determined by the first order Bragg condition in eq. (1). For each point of the RC the angle was rotated by 0.025° for the relatively narrow first order Bragg peak in reflectivity, and 0.03° for the wider TR RC spectra which range from 0° incidence to 2.5° to capture any structure in the tail end of this reflectivity peak. The angular resolution of 0.025° is near the expected lower limit of the sample goniometer. For the fits over all 62 Bragg RC spectra, and 88 TR RC spectra, the average $2p_{3/2}$ peak separations between Fe metal and Fe II+ were 2.5 eV, and between Fe metal and Fe III+ were 4.6 eV, which agrees well with expectations from the literature for metallic Fe, Fe II+, and Fe III+ [35]. The SW method in this instance uses the multilayer MgO/Fe superlattice to create an x-ray SW with a period matching the superlattice period of ~33 Å, as determined from the combined MgO, Fe, and FeO layer thickness from final results. This period was measured with hard x-ray reflectivity (XRR) measurements at 8,000 eV to be 38 Å (Figure 12), and from STEM measurements to be 37 Å (Figure 5), which is within the error range of the SW result and the nominal growth parameters.

The rocking curve results are shown in Figure 6. Each RC data point is the background-subtracted peak area for a spectrum measured at a single incidence angle $θ_i$ (see Figure 3). The Fe III+ rocking curve is flat and shows only noise which indicates that the Fe III+ is not present in a distinct layer. The Fe III+ RCs are thus not reported here [36].

Standard HXPS, such as shown in Figures 2-4, and modeled by the SESSA program, is sensitive to layers near the top surface. However, the phase modulation of the electric field in SW-HXPS allows for a more detailed view of the structure. Structural determination of the superlattice, shown in Figure 7, was determined from the experimental rocking curves by matching them to photoemission intensity predictions using known x-ray optical properties of the individual materials in the layers [37] with the YXRO program package [38]. Layer thicknesses and interdiffusion were allowed to vary, with the top layer independent of the other 8 repeats underneath. No strong fit to the phases of the RCs was found without varying this top bilayer independently. The other layers were not varied independently to prevent overfitting, after checking the lower layer consistency by the STEM results, which were within experimental error.

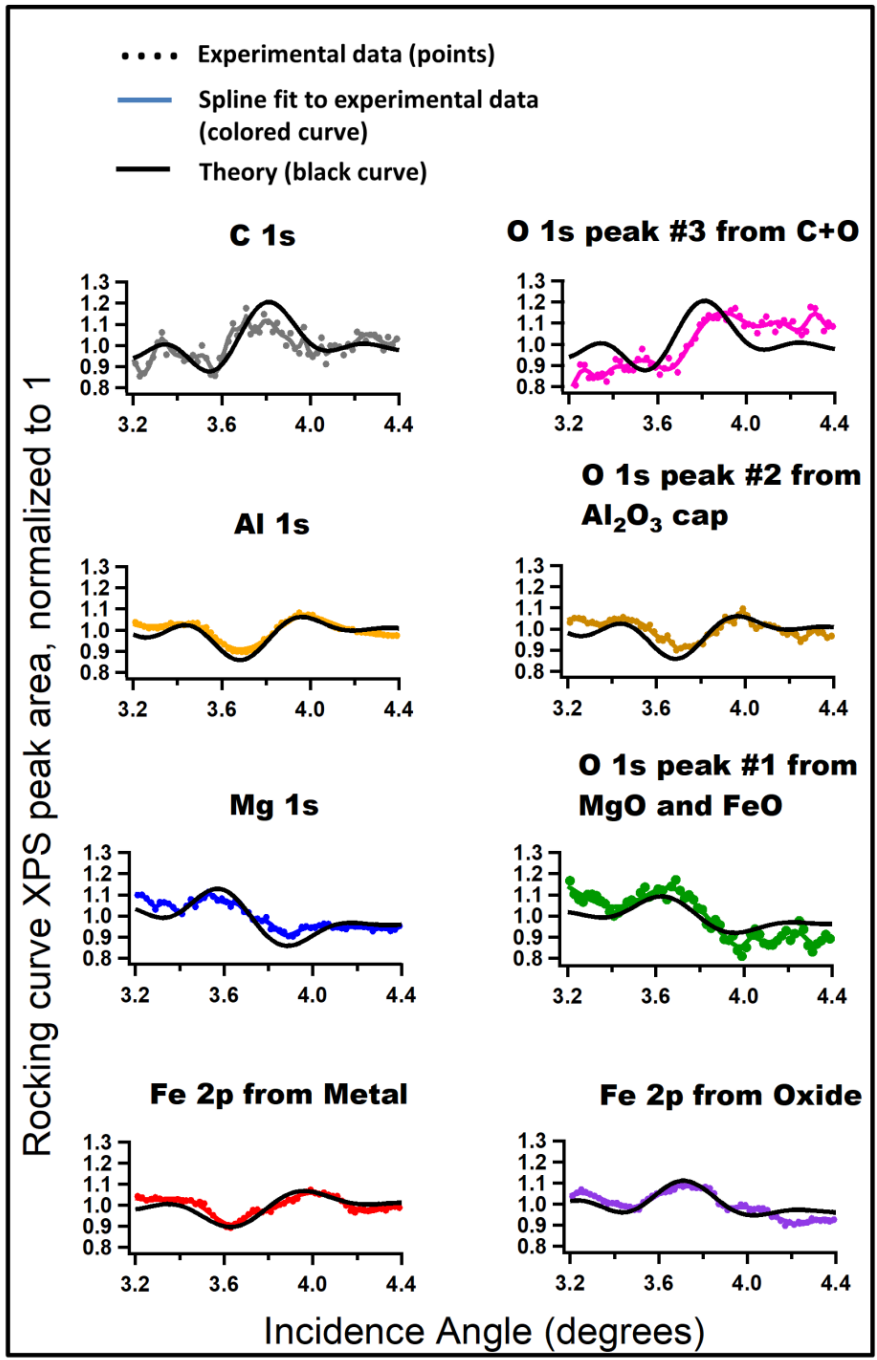

**Figure 6**: SW-HXPS RCs at a photon energy of 3,000 eV. Experimental RCs of fitted HXPS peak intensity (points) with theory fits from the YXRO (black lines) x-ray optics program. All RCs average intensity have been normalized to one in the low and high angle wings of the curves and a spline fit of the experimental data (colored lines) has been included for clarity. Fe 2p RCs also have a 9 point Savitzky-Golay smoothing applied. RC for Fe III+ not shown due to a lack of RC modulation which indicates that the Fe III+ is not present in a distinct layer [36].

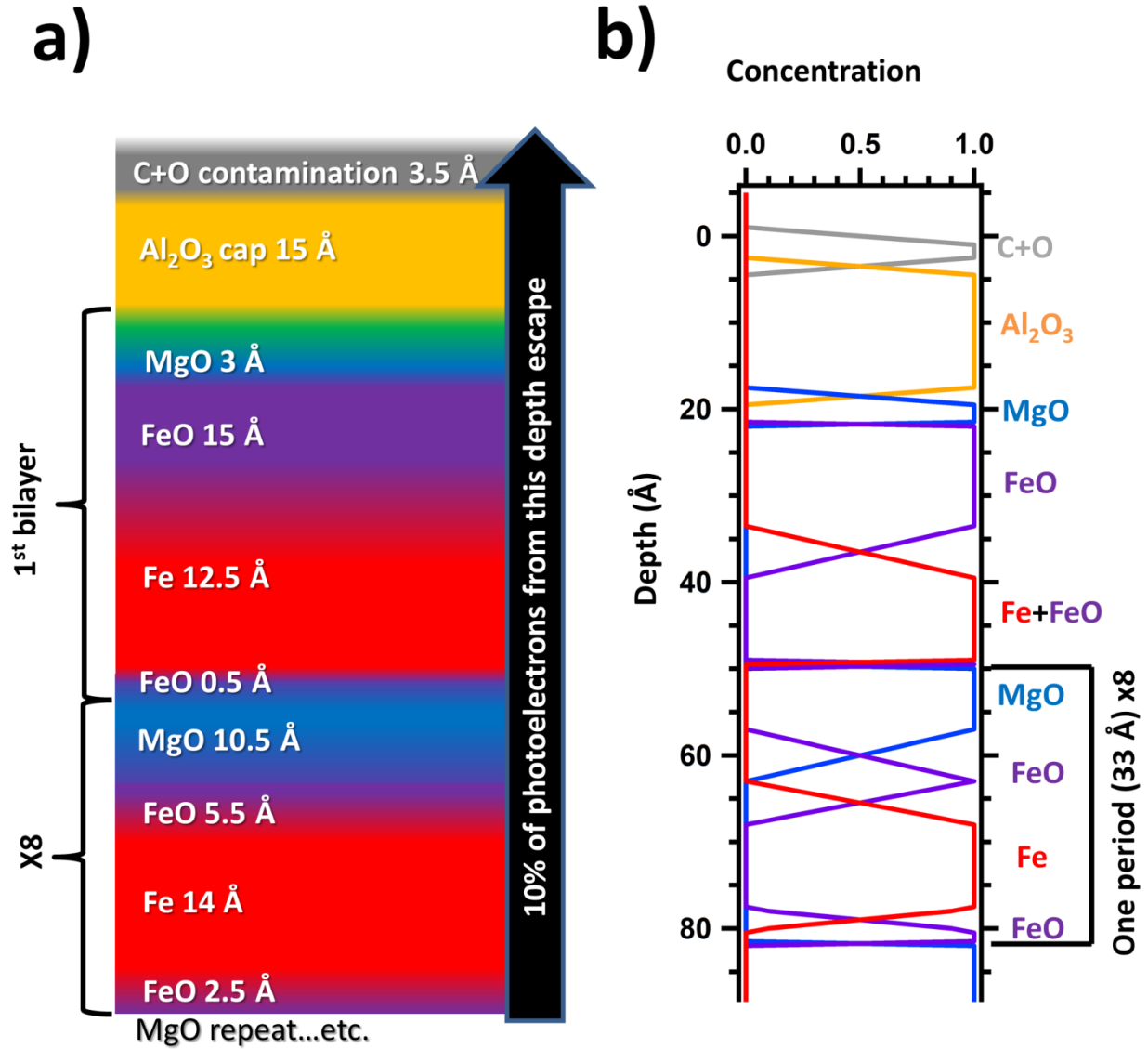

**Figure 7**: **a)** Structure determination from RCs and YXRO fits for the surface through the first two bilayers of the superlattice. Layer order, thickness, and interdiffusion of all other MgO, FeO, and Fe layers repeat as shown for the indicated bilayer. Black arrow shows depth in sample where 10% of excited Fe 2p photoelectrons originating from this location escape. **b)** Concentration from a) as a function of sample depth where 0 is the average surface location.

Fitting was done using a standard R-factor, least squares, comparison of experiment and theory, and also making use of a new, more rapid and accurate search algorithm based on a versatile Black Box Optimizer [39]. The total R factor is the sum of each least squares fit of the YXRO computed and normalized RC with the relevant normalized experimental RC. Additionally, each RC R factor in the sum is given a coefficient proportional to the counting statistics of the respective HXPS core level. For example, the C 1s and O 1s peak #3 core levels which are both low intensity have lower coefficients for their RC R factors. The Black Box Optimizer tests sample structures within a bounded range, which for this study was centered around the STEM and XRR results for the period of the sample bilayers. With each computed R factor the Optimizer program updates a predicted model of the many-dimensional R factor surface in the parameter space of the independent variables, which include layer thicknesses and interdiffusion lengths in this case. We used 10,000 sample calculations, and found a good surface estimation at around 7,000 calculations. The Black Box method utilized [39] avoids local minima that often arise with other fitting approaches, and has been found to speed up data analysis by 10-100 times.

Theoretical curve fits for core level RCs of all elements within the sample are shown in Figure 3 and Figure 4a. Resolution of the SW technique is estimated from a number of prior studies to be c.a. 1/10 of the standing wave period, around ±3-4 Å for this sample [40].

This sample structure was also used in a simulation of XPS photoemission intensity to compare to the full sample survey using the SESSA program [29] as a self-consistency check (normalized intensities are shown in Table I). This simulation uses the SW fitted results for the bottom 8 bilayers but varies the thicknesses in the top bilayer to fit the experimental values. Due to the exponential drop-off in photoemission intensity with depth, the HXPS measurements only access information in these top layers, whereas the superlattice mirror also affects the x-ray SW in the SW-HXPS measurements. As far as the top Mg 1s layer is concerned, the thicknesses for the $Al_2O_3$ and top FeO layers are overestimated at 2.3 and 1.9 times expectation from the SW-HXPS results, with the C+O contaminant layer overestimated by a factor of 2.9 and the topmost metallic Fe layer underestimated by a factor of 4. Since a more reasonable level of error in the traditional XPS estimation is around 10% [29], we believe that this indicates that the assumptions of the SESSA model we used were too restrictive. Specifically, SESSA does not allow for interdiffusion between layers and all layers were assumed to be continuous and consistent over the sample. Given these discrepancies between the SESSA and SW-HXPS estimations, we conclude that there is likely significant disruption to the continuity of this topmost bilayer. This is possibly due to inclusions of Fe (and Fe oxides) in the top MgO layer and increased buckling of the entire layer as compared to the lesser buckling of the lower layers (Figure 5, right), along with extreme roughness of the alumina capping layer as observed during preparation of the STEM sample.

The HXPS core level results, summarized in Figure 3 and Figure 4a, show information about the chemical species present in the sample, while the RCs in Figure 6 determine their depth profile in the sample. The shape of the RCs can give qualitative information about the sample as well. That is, phase matched rocking curves derive from chemical species with matching depth distributions through the sample; the Al 1s and O 1s peak #2 rocking curves show this quite clearly. Mg 1s and O 1s peak #1 have similar, but not exactly matching rocking curves. This arises from O peak #1 being a sum of O from both MgO and FeO, which have similar enough binding energies that they are not distinguishable in the O 1s core level peak (Figure 3). If they were separated enough in binding energy, then the O 1s core level would be able to be separated into four peaks with distinct rocking curves. As it is, separating the peak into three Voigt peaks creates three rocking curves with distinct phases, thus the O 1s core level has three distinct peaks rather than a single or double peak with an asymmetry or shoulder. In the case of asymmetries, shoulders, and satellite peaks these structures will have rocking curves mimicking the primary peak. In one additional case RCs can have matching phases, if the difference in sample depth is a multiple of the sample period $d_{ML}$, which is equivalent to a phase difference of $2\pi$. As a visual guide, see Figure 8, which shows the calculated squared electric field intensity inside the sample using the YXRO program and the structure from Figure 7. The vertical lines labeled 1 and 2 correspond to spectra at incidence angles corresponding to the minima and maxima of the RC for the very top of the topmost metallic Fe layer. The bottom of these lines show that the top of the next metallic Fe layer has RC phase minima and maxima 0.1° shifted compared to the topmost metallic Fe layer. These HXPS signals from different layers add up to stretch the measured RC along the incidence angle axis. Another case with similar but not matching rocking curves is from C 1s and O 1s peak #3. The C is from surface contamination, but the O 1s peak #3, with a binding energy matching several different potential contaminants such as CO and hydrolyzed O, does not closely follow the signature for the surface layer (as shown with the theory curve in Figure 6). The error for the thickness determination of this contaminant layer by RC analysis is expected to be greater than for other rocking curves measured in this sample due to the low intensity of the two photoelectron peaks involved and consequently greater noise in the rocking curve. This does suggest the possibility that some O at the binding energy of peak #3 could be present below this surface level, such as within the alumina capping layer, or between this layer and the top MgO layer. However, a quantitative determination is unreliable with the low counting statistics in our data.

The Fe core levels, in particular Fe 2p, show the presence of both metallic and oxidized Fe. Fe has two oxidation states which could be present in this sample, and a plethora of geometric orientations with respect to oxygen within those oxidations [41]. The Fe oxidation states that match with the binding energies of the oxidation peaks in the Fe 2p core levels (Figure 4b) are in the range for Fe II+ and Fe III+. The lineshapes for Fe II+ and Fe III+ have been shown to be predictable over many different Fe oxide crystal structures [35, 42]. We take advantage of this by using reference sample spectra of metallic Fe, Fe II+ (from FeO), and Fe III+ (from $Fe_2O_3$) to fit the Fe 2p spectra for the superlattice. An example fit is shown in Figure 4a. From this we find significant presence of Fe II+, and an indication of some presence of Fe III+. The lack of a rocking-curve signature at the Bragg peak for the very weak Fe III+ [36] suggests that it is not present in a coherent layer, as the rocking curves are visible even when dealing with very thin, singular layers. The relative percentages of the three fitted components for the averaged Fe 2p core level shown in Figure 4a are 36% for metallic Fe, 44% for Fe II+, and 20% for Fe III+. From the RC data it is clear that there is an Fe II+ component closer to the surface, and thus with signal enhanced, compared to the metallic Fe component. Fe II+ is present in both FeO and $Fe_3O_4$, but $Fe_3O_4$ contains as much Fe III+ as Fe II+ and thus is not likely present in any significant quantity in the sample as it does not match the stoichiometry of the results (Figure 4 and Figure 7). FeO, however, is known to have a common defect in the crystal structure leading to a small amount of Fe III+ [42]. The main oxidized Fe is likely FeO, with Fe III+ defects. From the rocking curve data (Figure 6) it is clear that this FeO is located at the interface of the Fe and Mg layers, predominantly at the top Fe interface where MgO is grown on Fe. These conclusions are supported by classic angle-resolved XPS measurements (ARXPS) measurements of the Fe 2p core level photoelectron emission angles of $\theta_e$ = 90°, 45°, and 30° at hv = 3,000 eV, shown in Figure 9a. Due to the increased surface sensitivity of the measurements at lower emission angles and lower excitation energies, where average depth of photoemission is proportional to

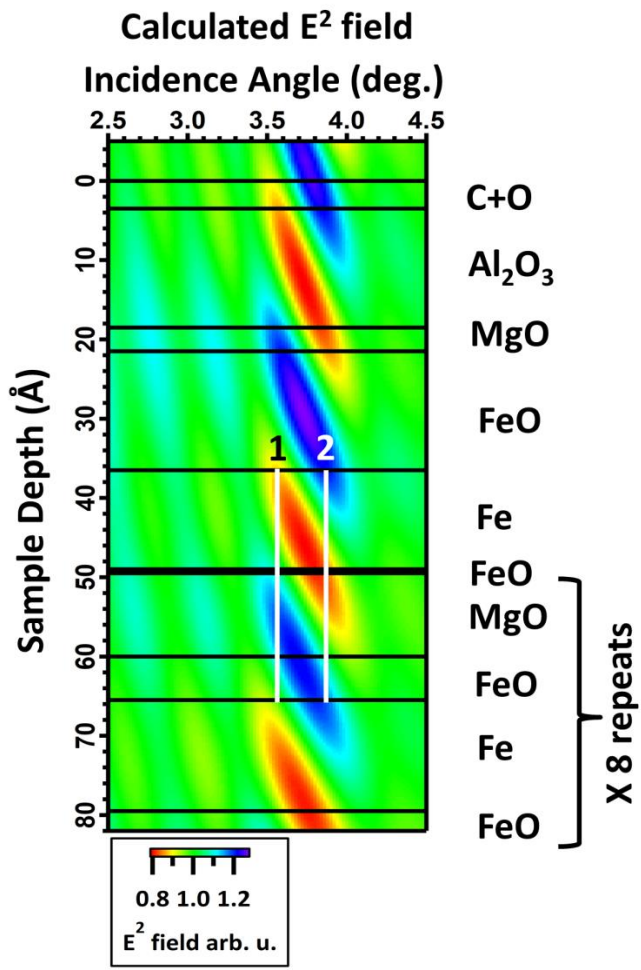

**Figure 8**: Electric field intensity profile calculated in the YXRO program, using hv = 3,000 eV and the final sample structure from figures 6, 10, and 11. The top 2 bilayers are shown, and the indicated layers at the bottom repeat to form the multilayer "mirror". This field was used for the YXRO HXPS RC calculations. The top of the vertical guidelines 1 and 2 indicate the minimum and maximum metallic Fe 2p RC intensities at the top of the topmost metallic Fe layer, respectively. The bottom of these guidelines show a phase shift of 0.1° for the metallic Fe 2p RC at the top of the second topmost metallic Fe layer.

sin($\theta_e$) [31], the enhanced component of these peaks at higher binding energies supports the conclusion that the oxidized component of the Fe is closer to the surface. Fitting the spectra in Figure 9a in the same way as the Fe 2p spectrum in Figure 4a, the metallic Fe contribution is unchanged at 36% from $\theta_e$ = 90° to $\theta_e$ = 45°, but decreases to 25% at $\theta_e$ = 30°. For the oxide components the Fe II+ component decreases with increased emission angle, from 44% to 41% at $\theta_e$ = 90° to $\theta_e$ = 30°. The Fe III+ component mirrors that with 20% at $\theta_e$ = 90° and 34% at $\theta_e$ = 30°.

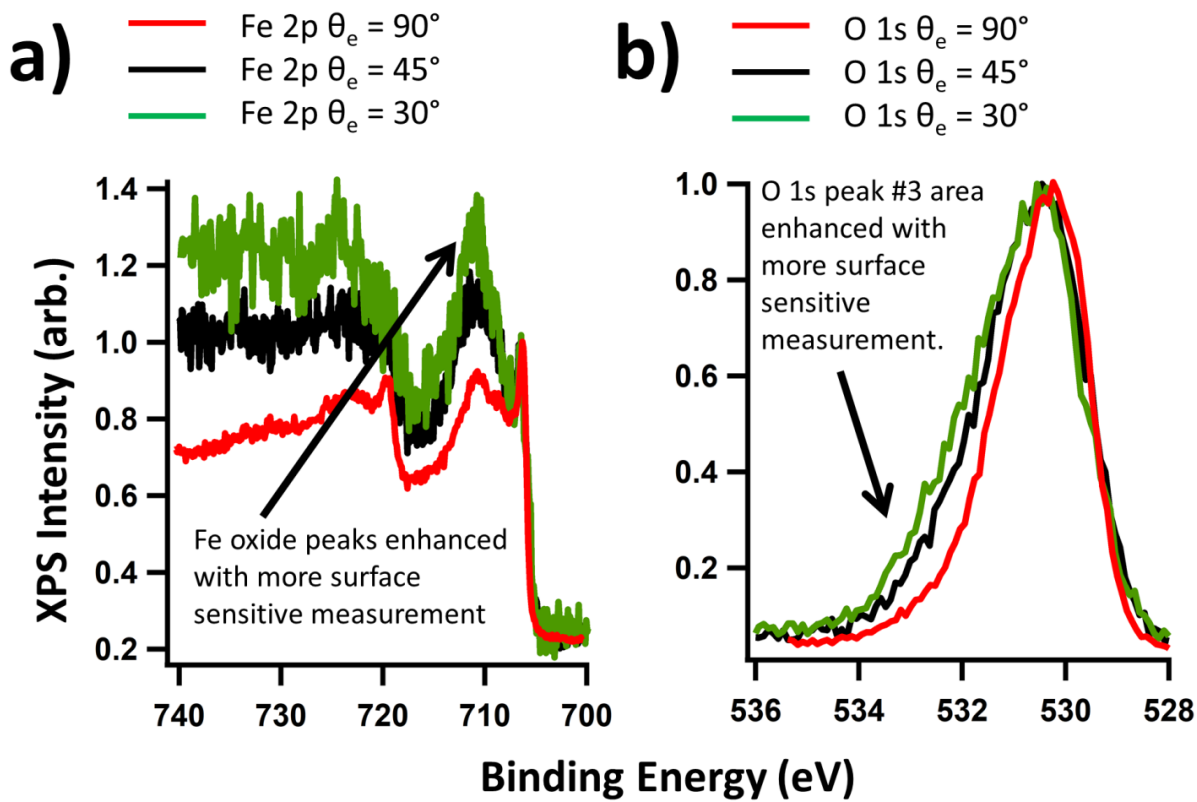

**Figure 9:** Comparison of **a)** Fe 2p and **b)** O 1s spectra measured at hv = 3,000 eV at emission angles of $\theta_e$ = 90, 45, and 30°. All peaks are normalized to metallic Fe and the highest intensity of O 1s.

Previous studies have shown that a completely clean Fe/MgO interface with no FeO bonding is unlikely, even in a sample with minimal interface roughness because of the presence of O on the surface lattice site in the MgO crystal with orientation of (001), and a preference for Fe to be located on the O sites [20]. Growth is favored for MgO on Fe [43], although in this case we observe significant oxidation for this interface, and only the expected intermixing at any interface with natural atomic steps for the Fe on MgO interface. With an experimental error of c.a. 3-4 Å the oxidation layers with thicknesses below this range indicate a presence of FeO at these interfaces, however our data does not permit quantitatively determining a layer thickness that is distinct from layer roughness, intermixing, or normal Fe to O bonding at this boundary.

The O 1s core levels also support the Fe 2p RC results for Fe oxidation (Figure 3). There are three separable peaks within the O 1s core level that each have different rocking curve phases, and thus are from distinct layers in the sample. Additional angle-resolve XPS data in Figure 9b shows intensity in the region of oxygen peak #3 is at least partially from atoms closer to the sample surface than the protective capping layer. As discussed above, the difference in phase between the oxygen peak #3 RC and that of C 1s indicates that either there is possibly a small amount of O matching this binding energy at another sample depth, or another O contaminant peak with a similar binding energy at concentrations too low to discern. Oxygen peak #2 (clearly associated with $Al_2O_3$ in the rocking curves, as shown by their matching phases) is then located just below that, and oxygen peak #1 (associated with both MgO and FeO) contains atoms occurring below that. Oxygen from the $Al_2O_3$ cap is 0.8 eV separated in binding energy from the O 1s peaks from MgO and FeO, and O from surface adsorption and contamination is 2.5 eV separated from MgO and FeO. The O 1s binding energies of MgO and FeO, however, are too closely positioned to resolve into separate peaks. As expected in a case with inseparable binding energies, the sum of the theoretical rocking curves for the O in FeO and MgO matches the RC of peak #1 better than either individual theoretical RC. Stoichiometrically, the amount of O at the interface of MgO on Fe growth matches with the presence of a fully formed FeO layer, whereas the lower interface of Fe on MgO is more consistent with O bonding to Fe at a rough interface, with interdiffusion on the scale of atomic steps, since the SW fitted FeO layer thickness is below that of a full unit cell (Figure 7).

The Al 1s peak and associated O 1s peak #2 show an overall well oxidized Al layer, as metallic Al was not present in the XPS core level results. The Mg 1s peak also shows no indication of the presence of additional Mg at a second binding energy, such as Mg not bonded to O (Figure 3).

Additional RC data in the total reflection (TR) region at 3,000 eV taken at beamline 9.3.1 of the ALS are shown in Figure 10 and Figure 11. The same parameters for the YXRO fits from the RC results of Figure 6 (shown in Figure 7) were used. The theory curves in Figure 11 (black curves) show good agreement with the RC fitting results (Figure 6). In the TR regime, at the lowest of grazing incidence angles, these measurements become increasingly surface sensitive as the incidence angle approaches zero incidence. Qualitatively, as the incidence angle increases, the first photoemission peaks to appear above background are expected to correlate to elements present closer to the sample surface. At the higher incidence angle range of the TR RCs, we also observe Kiessig fringes (Figure 10, especially Mg 1s, Figure 13). These arise from reflection from the top and bottom of the multilayer stack, and indicate the total thickness of the sample above the substrate, by the equation,

$$D_{total} = \frac{\lambda_{photon}}{2\Delta\theta_{Fringe}}, \quad (4)$$

where $D_{total}$ is the sample stack height, $\lambda_{photon}$ is the photon wavelength, and $\Delta\theta_{Fringe}$ is the separation of the fringe peaks. This simplified formula includes the assumption that the Bragg angle and critical angles of the experiment are small [44]. According to this analysis the multilayer is 395 Å thick, as compared to reflectivity data (see Figure 13) which indicates a bilayer thickness of 38 Å, or an estimated superlattice total thickness of 342 Å (both estimates not including the $Al_2O_3$ capping layer or surface contamination). Due to the lack of flat layers, and the elision of the effect of each material's optical properties in the estimate of eq. (4), the fitted sample structure is in reasonable agreement with the Kiessig fringe estimate.

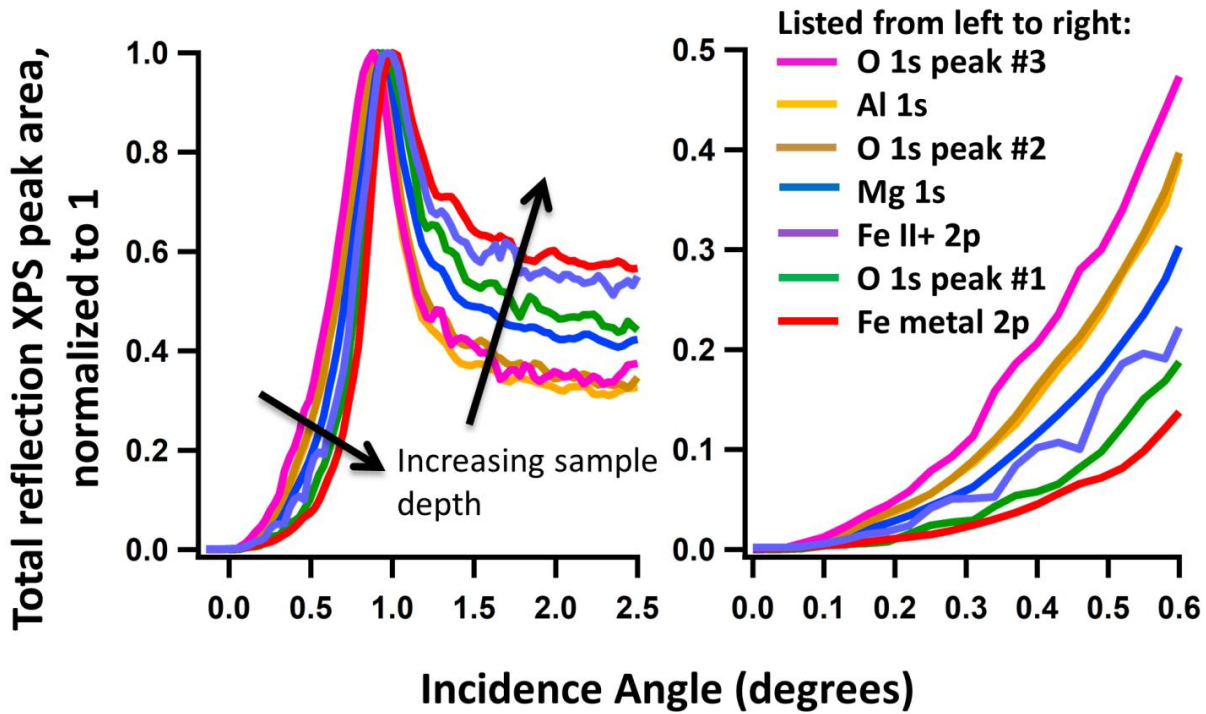

**Figure 10:** HXPS TR RCs at the photon energy of 3,000 eV. All RCs highest intensity have been normalized to 1. RC for Fe III+ not shown due to lack of RC modulation [36]. Typically elements that are closer to the surface of the sample have an onset at lower incidence angle, and are in reverse order at higher angles due to simple x-ray attenuation effects. At right is an enlargement of the low angle region of the data on the left. A 5 point Savitzky-Golay smoothing was applied to the Fe 2p curves. Note the oscillations near 0 incidence for Fe II+ 2p, which were too fine to appear in the theory simulations.

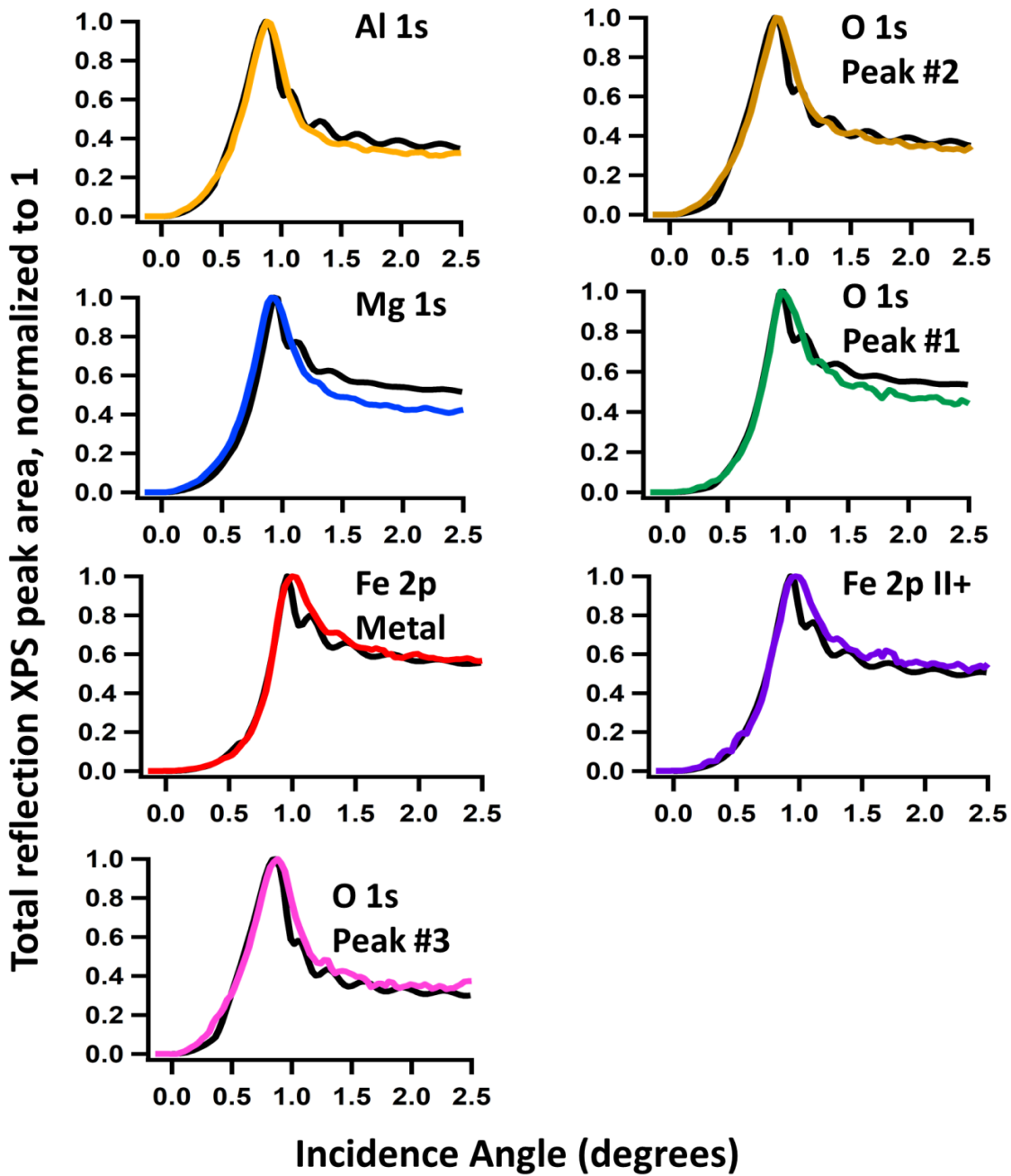

**Figure 11:** HXPS TR RCs at the photon energy of 3,000 eV, from figure 10. Colored curves are experimental data with YXRO theory fits shown in black. All RCs highest intensity have been normalized to 1.

HXPS is sensitive not only to the binding energy of different species, but to any local buildup of electric potential. In the depth-sensitive rocking curves this shows up as a shift in the binding energy position of the associated peak at discrete points in the rocking curve. None of the elemental rocking curves showed this shift, indicating no measurable build-up at any interfaces within the experimental resolution of 0.6 eV.

In a recent study by T. Thersleff et al. [25], a thick epitaxially grown Fe (50 nm single layer) on MgO (001) substrate exhibited no measureable FeO at the interface, and steps on the MgO surface were estimated to contribute to a total of 20 Å of mixing of Mg and Fe at the interface in TEM measurements [25]. Except for the very top layers of our sample, which showed much more intermixing, but still within the 20 Å range, the mixing we observed was under 10 Å, taking possible experimental error into account (see Figure 7). As noted above, though the fitting algorithm indicated FeO average layer thicknesses below the experimental error of 3-4 Å, these are interpreted to be on the order of a unit cell, approximately within the 4 Å range. In T. Thersleff et. al. [25], using the EMCD in EELS method, found an enhanced orbital-to-spin moment ratio within 24 Å, or the first several unit cells, of the interface. This study also used density functional theory (DFT) calculations to model the magnetism at this interface, and suggested that additional experimental information regarding the charge transfer at the interface could result in a clearer understanding of the mechanism leading to the orbital moment enhancement. While our study involves a stack of nanometer-scale alternating Fe and MgO layers with many interfaces and T. Thersleff et al. [25] studied a single interface with a thick Fe layer on thick MgO, it is of interest to note that we found no evidence of this charge build-up in the chemical shifts of binding energies. With TEM any interface roughness/interdiffusion will cause some smearing in the measurement depending on the sample thickness. Although in a SW study the measurement is averaged over the beam spot and detector aperture area [38], this averaging is a different mechanism than that in TEM. The SW study treats the interfaces and their roughness as a scattering plane, versus averaging through the thickness of a TEM sample. This suggests these two techniques can be complementary for samples depending on individual experimental challenges.

STEM directly shows the intermixing and wavy buckling at the interfaces, as seen in Figure 5. Based on row-by-row counts of atomic columns in the STEM, strain from lattice mismatch, which is visible as dislocations at the interfaces, is propagated through the sample without significant relaxation through planar defects within individual layers. That could contribute to the curvature of the layers, although as noted previously, and layer smearing complicates this determination. STEM indicates regularity in the layers through the sample, with a possible exception for the top layers, which have a thinner Fe layer. Although much of the shape of the RCs are somewhat sensitive to the entire sample stack, which determines the electric field profile within the sample, the total measured area of the photoelectron peaks are more sensitive to the top layers which produce the photoelectrons. The only way to reconcile the HXPS total peak area with the theoretical results is by treating the top MgO/Fe bilayer of the sample as being different from the 8 repeated layers below (see Figure 7). In STEM preparation the top alumina cap was found to have a very rough surface. Below this there is indication that the layers buckled into waves, and that there are possible inclusions indicated by locally brighter contrast in the top MgO layer that also could not be modeled as flat layers with interdiffusion (Figure 5). The HXPS measurements are an average over the beam spot and acceptance angle of the spectrometer, which for our experimental geometry, at very low incidence angle, and with the detector settings used, are averaging over millimeters of sample surface. This combined with the theory tools available in YXRO [38] and SESSA [29], which include perfectly flat layers with regular linear diffusion between them, restricts the exact representation of the model in this topmost layer. The average effect, however, is quantifiable, and the RCs and STEM both indicate that the bottom 8 bilayers are coherent and epitaxial in the STEM measurement, and well defined enough to produce an x-ray SW. Due to previous measurements by the sample growing team on similar samples with more bilayers than those in this study [13, 22], the differences found in this top layer could be restricted to this sample, or could be from the growth or coverage of the capping layer of alumina.

CONCLUSIONS:

We have thus shown that SW-HXPS can be a powerful tool for describing the chemical and electronic configurations at a very important, buried MTJ interface. The SW results indicate a thinner MgO layer than the nominal growth, 10.5 Å in the repeated layers versus 15 Å. The Fe repeated layers are similarly smaller. For a vertical expansion of 1:1.77 Å for the oxidation of Fe into FeO based on bulk values [45, 46], the SW results indicate an Fe thickness (unoxidized) of 19 Å in the repeated layers as compared to nominal value of 20 Å. In the SW data, roughness and inconsistencies in the layers is predominantly indicated by a wider, broader RC signature than expected, as seen in the SW results (Figure 6). The relative phases of the RCs result in an indication of the average layer thicknesses in the area of the beam spot, and we can see that a clear RC and TR RC signature was measured and could be fit with currently available theory models, as in the YXRO program [38].

The MgO layers are ordered, and SW-HXPS indicates a layer thickness of 2-3 unit cells of MgO, which is in range for a functional Fe/MgO/Fe TMR device [9, 10, 16]. An average Fe layer thickness of c.a. 5 unit cells is in line with predictions for maximized interfacial Fe magnetic moments [14], but indications of Fe lattice relaxation at these thicknesses is not clearly confirmed in the STEM results. FeO is present at the interfaces, particularly where MgO is grown on Fe, without any indication of Mg in MgO being stripped of O. The Mg 1s HXPS lineshape supports this observation, since O vacancies in the MgO would affect the Mg binding energy and produce a secondary peak or shoulder. With an interface where Fe is oxidized without migration of O out of the MgO layer, there is support in the literature for the possibility of ferromagnetic exchange at the interface [19], and the possibility that this oxidation, since it occurs on both MgO interfaces, may not depress the TMR [21]. With the asymmetry where the Fe grown on MgO interface shows less Fe oxidation, it would be of interest to study the magnetic properties at each of these interfaces individually, including SW MCD in photoemission, as applied previously to an amorphous Fe/MgO heterostructure [18].

From our SW-HXPS results, and self-consistent with additional multiple angle angle-resolved HXPS data, we find FeO at the Fe/MgO interfaces in the sample, with asymmetry where the majority of the oxidation is at the MgO on Fe interface. The sample is composed of otherwise well-defined, consistent, and epitaxial layers of metallic Fe and MgO, with expected dislocations at the interfaces from bulk lattice mismatch contributing to roughness, and some layer buckling. We note that the top sample bilayer of the sample stack was found to have significant differences compared to the other layers, with evidence of possible Fe inclusions in the top

MgO layer. We note that the O with the highest binding energy is mostly associated with the surface C layer, but due to the weakness of the peak the source of the minor phase mismatch with the C 1s RC is undetermined.

The SW-HXPS and STEM results are consistent within error for the bilayer thickness and layer intermixing. The bilayer thicknesses of 37 Å (STEM) and 33 Å (SW-HXPS) are in line with the 3-4 Å experimental of the SW-HXPS fitting error. This study thus confirms the structural conclusions of parallel STEM measurements on the sample, and demonstrates for another interesting spintronic system the utility of SW-HXPS for the quantitative study of multilayer heterostructures. These results suggest further study, including MCD measurements sensitive to the interfacial magnetism, where the individual interfaces of MgO on Fe and Fe on MgO can be analyzed separately and locally. The broad applicability of SW-HXPS to studying spintronic heterostructures is also demonstrated. The sampling of the literature on Fe/MgO based MTJ systems summarized in this paper demonstrate the vital importance of the interface structure, including oxidations states and stoichiometry, to the magnetic properties, as well as the sensitivity of the final structures to growth techniques and environments. We conclude that a careful, deep probing, non-destructive measurement technique that is sensitive to local charge and electronic configurations such as SW-HXPS is important for new developments in Fe/MgO MTJ structures.

**Acknowledgements:**

C.S.C. would like to acknowledge discussions with S. Ueda and G. Castro and use of their HXPS Fe and Fe oxide data for fitting and comparisons.

C.S.F. has been supported for salary by the Director, Office of Science, Office of Basic Energy Sciences (BSE), Materials Sciences and Engineering (MSE) Division, of the U.S. Department of Energy under Contract No. DE-AC02-05CH11231, through the Laboratory Directed Research and Development Program of Lawrence Berkeley National Laboratory, through a DOE BES MSE grant at the University of California Davis from the X-ray Scattering Program under Contract DESC0014697 (for experiments at the Advanced Light Source and for travel support to carry out experiments at Soleil), and through the APTCOM Project, "Laboratoire d'Excellence Physics

Atom Light Matter" (LabEx PALM) overseen by the French National Research Agency (ANR) as part of the "Investissements d'Avenir" program. C.S.C. has been supported by the GAANN and Dissertation Year Fellowship programs of UC Davis. Graduate student researcher-work study and departmental fellowship was awarded for A.K. at UC Davis and A.R. was funded by the Royal Thai Government. STEM measurements at the Molecular Foundry were supported by the Office of Science, Office of Basic Energy Sciences, of the U.S. Department of Energy under Contract No. DE-AC02-05CH11231.

**Appendix:**

X-ray reflectivity in the hard x-ray energy range of 8,000 eV was measured as an independent indicator of the first order Bragg peak in a system with precisely calibrated energy source and goniometer. Results indicate a 38 Å bilayer as estimated using a ratio of 1.3:1 Fe:MgO from grower specifications and estimation from STEM measurement, and an incidence angle of 1.24° from the x-ray reflectivity measurement at 8,000 eV as shown in Figure 12. Using eq. (1) this results in an estimation of the bilayer thickness of 36 Å. Fitting the reflectivity measurement with an online multilayer reflectivity model available online from the Center for X-Ray Optics at LBNL [47], and plotted in blue in Figure 12, this period is estimated at 38 Å.

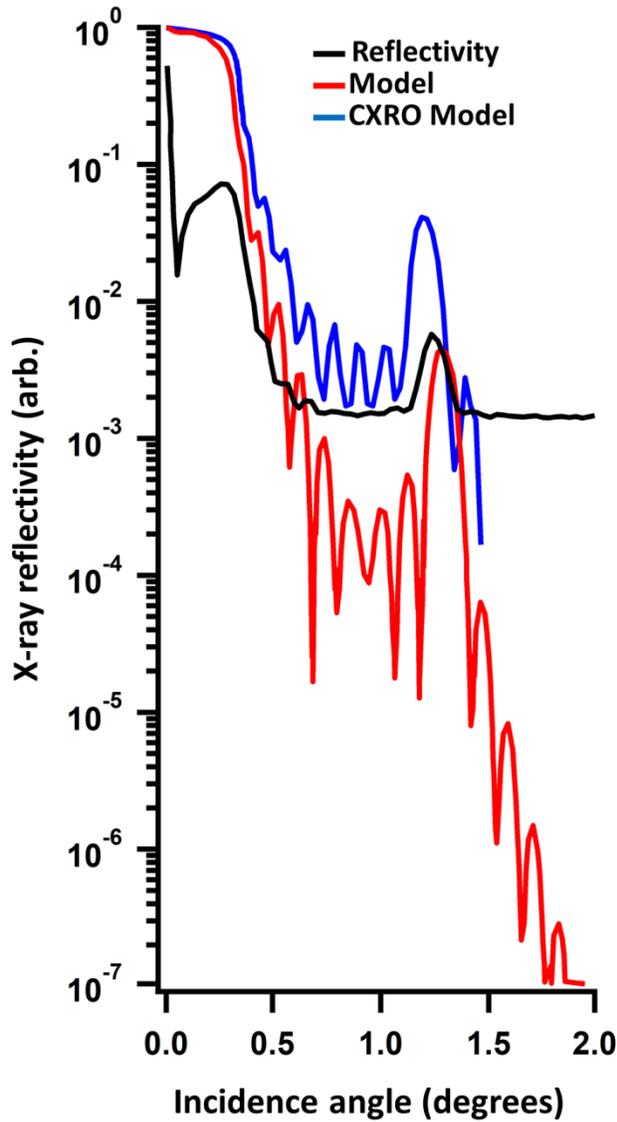

**Figure 12:** X-ray reflectivity at hv = 8,000 eV (black curve) and model reflectivity (red curve). Due to a small sample size compared to the beam, the reflectivity may be underestimated at 0.57%. Angle of first order reflection is at 1.24°. Using the model of a 9 layer stack with Fe layer 1.3 times the MgO layer thickness [47] results in a calculated bilayer thickness of 38 Å (blue curve).

Additionally, in regions outside of the high-reflectivity incidence angles, there is another periodic variation that can be seen in the RCs. This is most visible in the high incidence angle tail-end of the TR RCs, as shown in Figure 13. This figure shows an enlarged view of this variation, called Kiessig fringes [44], where the sloped background is removed. Using eq. (4), this gives a rough estimate of 395 Å for the thickness of the full layer stack of the sample, from the substrate to a high-contrast index of refraction interface at the top.

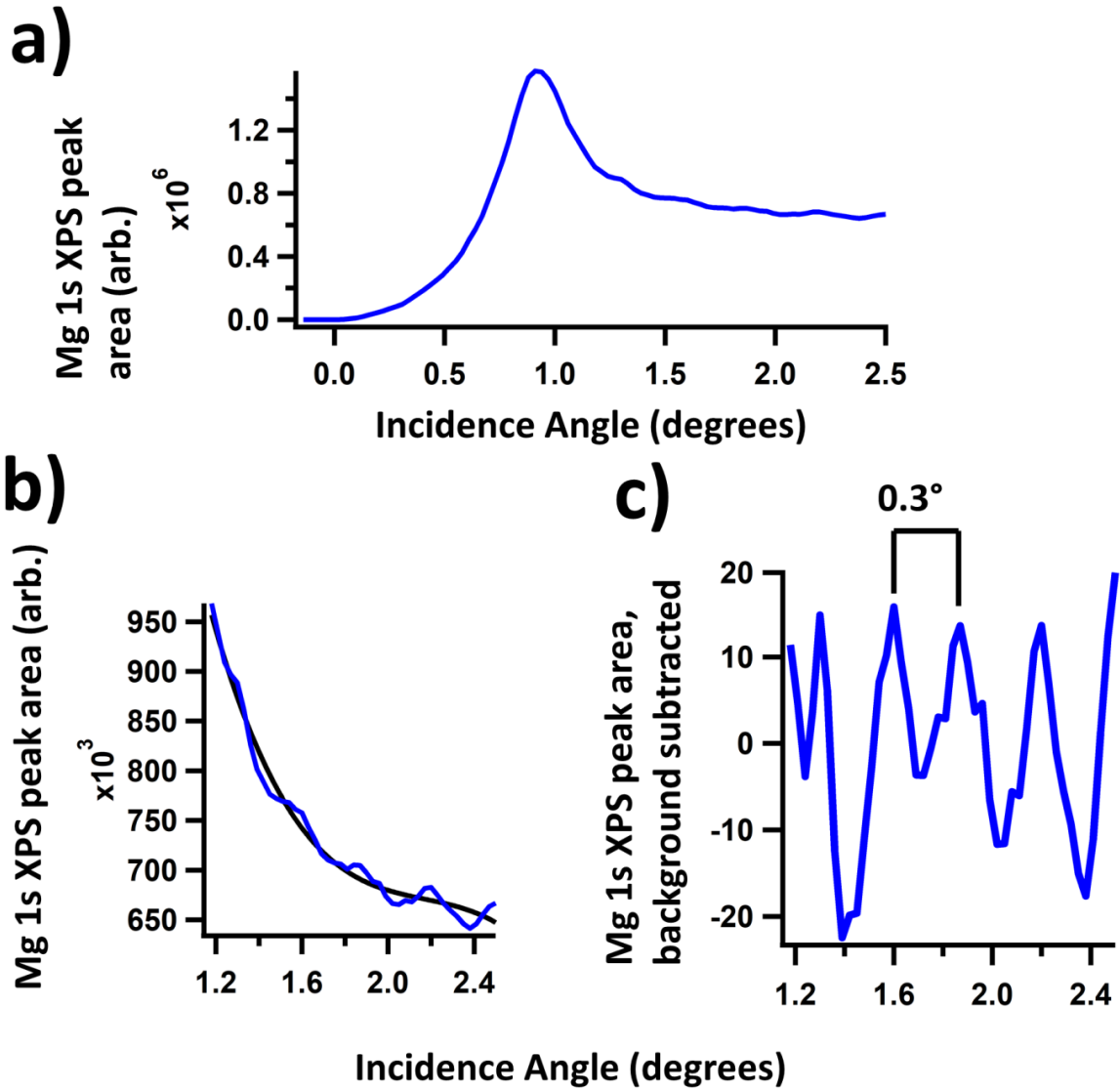

**Figure 13: a)** TR RC of Mg 1s. **b)** Higher incidence angle portion of Mg 1s TR RC from a) (blue curve). Periodic variations are the Kiessig fringes from reflections between the substrate and surface of the superlattice. Polynomial fit of Kiessig fringe region is shown (black curve). **c)** Kiessig fringes from high incidence angle TR RC of Mg 1s with polynomial background from b) subtracted.

47 CXRO multilayer reflectivity: http://henke.lbl.gov/optical_constants/multi2.html accessed: Oct. 5, 2018.

37